\documentclass[aps,prd,twocolumn,groupedaddress,showpacs,nofootinbib]{revtex4}
\usepackage{graphicx}

\begin{document}

\title{Constraining dark energy models using the lookback time to galaxy clusters and the age of the universe}

\author{S. Capozziello}
\author{V.F. Cardone}
\thanks{Corresponding author, email: {\tt winny@na.infn.it}}
\affiliation{Dipartimento di Fisica ``E.R. Caianiello'', Universit\`{a} di Salerno and INFN, Sez. di Napoli, Gruppo Coll. di Salerno, via S. Allende, 84081 - Baronissi (Salerno), Italy}
\author{M. Funaro}
\affiliation{Dipartimento di Matematica e Informatica, Universit\`{a} di Salerno, Via Ponte don Melillo, 84084 - Fisciano (Salerno), Italy}
\author{S. Andreon}
\affiliation{INAF-Osservatorio Astronomico di Brera, Via Brera 28, 20121 - Milano, Italy}

\begin{abstract}

An impressive amount of different astrophysical data converges towards the picture of a spatially flat universe undergoing a today phase of accelerated expansion. The nature of the dark energy dominating the energy content of the universe is still unknown and a lot of different scenarios are viable candidates to explain cosmic acceleration. Most of the methods employed to test these cosmological models are essentially based on distance measurements to a particular class of objects. A different method, based on the lookback time to galaxy clusters and the age of the universe, is used here. In particular, we constrain the characterizing parameters of  three classes of dark energy cosmological models to see whether they are in agreement with this kind of data,  based on time measurements rather than distance observations.

\end{abstract}

\pacs{98.80.-k, 98.80.Es., 98.65.Cw}

\maketitle

\section{Introduction}

The increasing bulk of data that have been accumulated in the last few years have paved the way to the emergence of a new standard cosmological model usually referred to as the {\it concordance model}. The Hubble diagram of Type Ia Supernovae (hereafter SNeIa), measured by both the Supernova Cosmology Project \cite{SCP} and the High\,-\,z Team \cite{HZT} up to redshift $z \sim 1$, first indicated that the universe is undergoing a phase of accelerated expansion. On the other hand, balloon born experiments, such as BOOMERanG \cite{Boomerang} and MAXIMA \cite{Maxima}, determined the location of the first and second peak in the anisotropy spectrum of  cosmic microwave background radiation (CMBR) strongly indicating that the geometry of the universe is spatially flat. If combined with constraints coming from galaxy clusters on the matter density parameter $\Omega_M$, these data indicate that the universe is dominated by a non-clustered fluid with negative pressure generically dubbed {\it dark energy} which is able, eventually, both to close the universe and drive its accelerated expansion. This picture has been further strengthened by the more precise measurements of the CMBR spectrum, due to the WMAP experiment \cite{WMAP}, and by the extension of the SNeIa Hubble diagram to $z > 1$ \cite{Riess04}.

After the discovery of this scenario, an overwhelming flood of papers, presenting a great variety of models which try to explain this phenomenon has appeared; in any case,  the simplest explanation is claiming for the well known cosmological constant $\Lambda$ \cite{LCDMrev}. Although the best fit to most of the available astrophysical data \cite{WMAP}, the $\Lambda$CDM model failed in explaining why the inferred value of $\Lambda$ is so tiny (120 orders of magnitude lower) compared to the typical vacuum energy values predicted by particle physics and why its energy density is comparable to the matter density right now (the so called {\it coincidence problem}). As a tentative solution, many authors have replaced the cosmological constant with a scalar field rolling down its potential and giving rise to the model now referred to as {\it quintessence} \cite{QuintFirst}. Even if successful in fitting the data, the quintessence approach to dark energy is still plagued by the coincidence problem since the dark energy and matter densities evolve differently and reach comparable values for a very limited portion of the universe evolution that is taken to coincide with the present era. Moreover, it is not clear where this scalar field originates from thus leaving a great uncertainty on the choice of the scalar field potential (see \cite{QuintRev} for comprehensive reviews).

The subtle and elusive nature of  dark energy has led many authors to look for completely different scenarios able to give a quintessential behavior without the need of exotic components.

To this aim, it is worth stressing that the acceleration of the universe only claims for a negative pressure dominant component, but does not tell anything about the nature and the number of cosmic fluids filling the universe. This consideration suggests that it could be possible to explain the accelerated expansion by introducing a single cosmic fluid with an  equation of state causing it to act like dark matter at high densities and dark energy at low densities. An attractive feature of these models, usually referred to as {\it Unified Dark Energy} (UDE) or {\it Unified Dark Matter} (UDM) models, is that such an approach naturally solves, al least phenomenologically, the coincidence problem. Some interesting examples are the generalized Chaplygin gas \cite{Chaplygin}, the tachyon field \cite{tachyon} and the condensate cosmology \cite{Bassett}.

A different class of UDE models has been recently proposed by some of us \cite{Hobbit} where a single fluid is considered whose energy density scales with the redshift in such a way that the radiation dominated era, the matter domination era and the accelerating phase can be naturally achieved. It is worth noting that such a model is extremely versatile since it can be interpreted both in the framework of UDE models and as a two-fluid scenario with dark matter and scalar field dark energy.

Actually, there is still a different way to face the problem of cosmic acceleration. As  stressed in Lue et al. \cite{LSS03}, it is possible that the observed acceleration is not the manifestation of another ingredient in the cosmic pie, but rather the first signal of a breakdown of our understanding of the laws of gravitation. From this point of view, it is thus tempting to modify the Friedmann equations to see whether it is possible to fit the astrophysical data with a model comprising only the standard matter. Interesting examples of this kind are the Cardassian expansion \cite{Cardassian} and the DGP gravity \cite{DGP}.

Moving in this same framework, it is possible to find alternative schemes where a quintessential behavior is obtained by simply taking into account the effective contribution to cosmology of some usually neglected fundamental physics \cite{capozcurv,curvature,review}. For instance, a cosmological constant term may be recovered as a consequence of a non\,-\,vanishing torsion field thus leading to a model which is consistent with both SNeIa Hubble diagram and Sunyaev\,-\,Zel'dovich data coming from clusters of galaxies \cite{torsion}. SNeIa data could also be efficiently fitted including higher-order curvature invariants in the gravity Lagrangian \cite{curvfit}. It is worth noting that these alternative schemes provide naturally a cosmological component with negative pressure whose origin is simply related to the geometry of the universe thus overcoming the problems linked to the physical significance of the scalar field.

It is evident, from this short overview, the high number of cosmological models which are viable candidates to explain the observed accelerated expansion. This abundance of models is somehow a result of having only a limited number of cosmological tests to discriminate among rival theories so that any new observable which could be added to the usual ones is welcome. To this aim, it is useful to remember that both the SNeIa Hubble diagram and the recently proposed angular size\,-\,redshift relation of compact radio sources \cite{AngTest} are distance based methods to probe cosmological models. From this point of view, it is  interesting to look for  tests based on time-dependent observables. To this aim, one can take into account the {\it lookback time} to distant objects since this quantity can discriminate among different cosmological models. The lookback time is observationally estimated as the difference between the present day age of the universe and the age of a given object at redshift $z$. Such an estimate is possible if the object is a galaxy observed in more than one photometric band since its color is determined by its age as a consequence of stellar evolution. It is thus possible to get an estimate of the galaxy age by measuring its magnitude in different bands and then using stellar evolutionary codes to choose the model that reproduces the observed colors at best. A quite similar approach was pursued by Lima \& Alcaniz \cite{LA00} who used the age (rather than the lookback time) of old high redshift galaxies to constrain the dark energy equation of state (see also \cite{Jimenez}). The same method was then applied also to braneworld models \cite{AJD02} and the Chaplygin gas \cite{AJD03}.

It is worth noting, however, that the estimate of the age of a single galaxy may be affected by systematic errors which are difficult to control. Actually, this problem can be overcome by considering a sample of galaxies belonging to the same cluster. In this way, by averaging the estimates of all galaxies, one obtains an estimate of the cluster age and reduces the systematic errors. Such a method was first proposed by Dalal et al. \cite{DAJM01} and then used by Ferreras et al. \cite{FMT03} to test a class of models where a scalar field is coupled with the matter term giving rise to a particular quintessence scheme. We improve here this analysis using a different cluster sample and testing three different approaches to the dark energy problem. Moreover, we add a further constraint to better test the dark energy models and assume that the age of the universe for each model is in agreement with recent estimates. Note that this is not equivalent to the lookback time as we will discuss below.

The layout of the paper is the  following. In Sect.\,II, we briefly present the three classes of cosmological models considered giving also the main quantities which we  need for lookback time test. The method  is developed in Sect.\,III and the used data are presented in Sect.\,IV. A detailed discussion of the results is the subject of Sect.\,V. We summarize and conclude in Sect.\,VI.

\section{Dark energy models}

As overviewed in the Introduction, many rival theories have been advocated to fit the observed accelerated expansion and to solve the puzzle of the nature of the dark energy. Actually, we may divide the different cosmological models in three wide classes. According to the models of the first class, the dark energy is new ingredient of the cosmic Hubble flow, the simplest case being the $\Lambda$CDM scenario and its generalization which we will refer to as QCDM model. This is in sharp contrast with the assumption of UDE models (the second class) where there is a single fluid described by an equation of state comprehensive of all regimes of cosmic evolution such as the one proposed in \cite{Hobbit} which we will consider here referring to it as the {\it parametric density model}. Finally, according to the third class models, accelerated expansion is the first evidence of a breakdown of the Einstein General Relativity (and thus of the Friedmann equations) which has to be considered as a particular case of a more general theory of gravity. As an example of this kind of models, we will consider the $f(R)$\,-\,gravity \cite{capozcurv,curvature,review,curvfit}. Far from being exhaustive, considering these three models allow to explore very different scenarios proposed to explain the observed cosmic acceleration.

In the following we will sketch these three dark energy models and derive all the quantities we will need for the analysis developed in Sect.\,III.

\subsection{The QCDM model}

Cosmological constant $\Lambda$ has now become a textbook candidate to drive the accelerated expansion of the spatially flat universe. Despite its {\it conceptual} problems, the $\Lambda$CDM model turns out to be the best fit to a combined analysis of completely different astrophysical data ranging from SNeIa to CMBR anisotropy spectrum and galaxy clustering \cite{WMAP,SDSS03,VSA}. As a simple generalization, one may consider the QCDM scenario in which the barotropic factor $w \equiv p/\rho$ takes a negative value with $w = -1$ corresponding to the standard cosmological constant. How such a negative pressure fluid drives the cosmic acceleration may be easily understood by looking at the Friedmann equations\,:

\begin{equation}
H^2 \equiv \left ( \frac{\dot{a}}{a} \right )^2 = \frac{8 \pi G}{3} (\rho_{M} + \rho_Q) \ ,
\label{eq: fried1}
\end{equation}

\begin{equation}
2 \frac{\ddot{a}}{a} + H^2 = - 8 \pi G p_Q = - 8 \pi G w \rho_Q \ ,
\label{eq: fried2}
\end{equation}
where the dot denotes differentiation with respect to cosmic time $t$, $H$ is the Hubble parameter and we have assumed that the universe is spatially flat as suggested by the position of the first peak in the CMBR anisotropy spectrum \cite{Boomerang,Maxima,WMAP}. From the continuity equation, $\dot{\rho} + 3 H (\rho + p) = 0$, we get for the $i$\,-\,th fluid with $p_i = w_i \rho_i$\,:

\begin{equation}
\Omega_i = \Omega_{i,0} a^{-3 (1 + w_i)} = \Omega_{i,0} (1 + z)^{3 (1 + w_i)} \ ,
\label{eq: omegavsz}
\end{equation}
where $z \equiv 1/a - 1$ is the redshift, $\Omega_i = \rho_i/\rho_{crit}$ is the density parameter for the $i$\,-\,th fluid in terms of the critical density $\rho_{crit} = 3H_0^2/8\pi G$ and, hereafter, we label all the quantities evaluated today with a subscript $0$. Inserting this result into Eq.(\ref{eq: fried1}), we get\,:

\begin{equation}
H(z) = H_0 \sqrt{\Omega_{M,0} (1 + z)^3 + \Omega_{Q,0} (1 + z)^{3 (1 + w)}} \ .
\label{eq: hvsz}
\end{equation}
Using Eqs.(\ref{eq: fried1}), (\ref{eq: fried2}) and the definition of the deceleration parameter $q \equiv - a \ddot{a}/\dot{a}^2$, one finds\,:

\begin{equation}
q_0 = \frac{1}{2} + \frac{3}{2} w (1 - \Omega_{M,0}) \ .
\label{eq: qlambda}
\end{equation}
The SNeIa Hubble diagram, the large scale galaxy clustering and the CMBR anisotropy spectrum can all be fitted by the $\Lambda$CDM model with $(\Omega_{M,0}, \Omega_Q) \simeq (0.3, 0.7)$ thus giving $q_0 \simeq -0.55$, i.e. the universe turns out to be in an accelerated expansion phase. The simplicity of the model and its capability of fitting the most of the data are the reasons why the $\Lambda$CDM scenario is the leading candidate to explain the dark energy cosmology. Nonetheless, we will first consider its generalization, the QCDM model, before concentrating on the $\Lambda$CDM scenario.

\subsection{The parametric density model}

In the framework of UDE models, it has been recently proposed \cite{Hobbit} a phenomenological class of models by introducing a single fluid\footnote{It is worth stressing that this model may be interpreted not only as comprising a single fluid with an exotic equation of state, but also as made of dark matter and scalar field dark energy or in the framework of modified Friedmann equations. Here, we prefer the UDE interpretation even if the results do not depend on this assumption.} with energy density\,:

\begin{equation}
\rho(a) = A_{norm} \left ( 1 + \frac{s}{a} \right )^{\beta - \alpha} \ \left [ 1 + \left ( \frac{b}{a} \right )^{\alpha} \right ]
\label{eq: rhor}
\end{equation}
with $0 < \alpha < \beta$, $s$ and $b$ (with $s < b$) two scaling factors and $A_{norm}$ a normalization constant. For several applications, it is useful to rewrite the energy density as a function of the redshift $z$. Replacing $a = (1 + z)^{-1}$ in Eq.(\ref{eq: rhor}), we get\,:

\begin{equation}
\rho(z) = A_{norm} \ \left ( 1 + \frac{1 + z}{1 + z_s} \right )^{\beta - \alpha} \
\left [ 1 + \left ( \frac{1 + z}{1 + z_b} \right )^{\alpha} \right ]
\label{eq: rhoz}
\end{equation}
having defined $z_s = 1/s - 1$ and $z_b = 1/b - 1$.  It is easy to show that $\rho \propto a^{-\beta}$ for $a << s$, $\rho \propto a^{-\alpha}$ for $s << a << b$ and $\rho \propto const.$ for $a >> b$. By setting $(\alpha, \beta) = (3, 4)$ the energy density smoothly interpolates from a radiation dominated phase to a matter dominated period finally approaching a de Sitter state. The normalization constant $A_{norm}$ may be estimated by inserting Eq.(\ref{eq: rhor}) into Eq.(\ref{eq: fried1}) and evaluating the result today. This gives\,:

\begin{equation}
A_{norm} = \frac{\rho_{crit,0}}{(1 + s)^{\beta - \alpha} \ (1 + b^{\alpha})} \ .
\label{eq: norm}
\end{equation}
The continuity equation may be recast in a form that allows us to compute the pressure for our fluid and then the barotropic factor $w = p/\rho$ obtaining \cite{Hobbit}\,:

\begin{equation}
w = \frac{[ (\alpha - 3) a + (\beta - 3) s] b^{\alpha} - [3 (a + s) + (\alpha - \beta) s ] a^{\alpha}}{3 (a + s) (a^{\alpha} + b^{\alpha})}
\label{eq: wr}
\end{equation}
which shows that the barotropic factor strongly depends on the scale factor (and hence on the redshift). Combining the Friedmann equations, we get for the deceleration parameter $q = (1 + 3 w)/2$ which, in our case, gives\,:

\begin{equation}
q = \frac{[(\alpha - 2) a + (\beta - 2) s] b^{\alpha} - [2 (a + s) + (\alpha - \beta) s] a^{\alpha}}
{2 (a + s) (a^{\alpha} + b^{\alpha})} \ ;
\label{eq: qr}
\end{equation}
inserting $a = 1$ gives the present day value as\,:

\begin{equation}
q_0 = \frac{(y - 1) \alpha + z_s [\alpha \ y - 2 (1 + y)] + (\beta - 4) (1 + y)}{2 (2 + z_s) (1 + y)}
\label{eq: qz}
\end{equation}
with $y = (1 + z_b)^{-\alpha}$.  Some straightforward considerations allows to derive the following constraints on $q_0$ \cite{Hobbit}\,:

\begin{equation}
\frac{1}{2} \left [ \frac{\beta - \alpha}{2 + z_s} - 2 \right ] \le q_0
\le \frac{1}{2} \left [ \frac{\alpha z_s + 2 \beta}{2 (2 + z_s)} - 2 \right ] \ .
\label{eq: qlimits}
\end{equation}
It is convenient to solve Eq.(\ref{eq: qz}) with respect to $z_b$ in order to express this one as a function of $q_0$ and $z_s$. It is\,:

\begin{equation}
z_b = \left [ \frac{\alpha (1 + z_s) + \beta - (2 + z_s) (2 q_0 + 2)}{\alpha - \beta + (2 + z_s) (2 q_0 + 2)} \right ]^{1/\alpha} - 1 \ .
\label{eq: solvezb}
\end{equation}
The parametric density model is fully characterized by five parameters which are chosen to be the two asymptotic slopes $(\alpha, \beta)$, the present day values of the deceleration parameter and of the Hubble constant $(q_0, H_0)$ and the scaling redshift $z_s$. As in \cite{Hobbit}, we will set $(\alpha, \beta) = (3, 4)$ and $z_s = 3454$ so that $(q_0, H_0)$ will be the two parameters to be constrained by the data.

\subsection{Curvature quintessence}

There is no {\it a priori} reason to restrict the gravity Lagrangian to a linear function of Ricci scalar $R$ as in the usual Einstein general relativity. Actually, higher order curvature invariants are introduced in quantum gravity theories so that it is worth considering the effect of such generalizations on the late evolution of the universe.

Replacing the Ricci scalar $R$ with a (up to now) generic function $f(R)$ of Ricci scalar in the gravity lagrangian, the resulting field equations may still be recast in the Friedmann\,-\,like form provided that the total energy density and the total pressure are written as\,:

\begin{displaymath}
\rho_{tot} = \rho_M + \rho_{curv} \ \ \ \ , \ \ \ \ p_{tot} = p_M + p_{curv}
\end{displaymath}
where the energy density and the pressure due to the higher\,-\,order curvature invariants are  \cite{capozcurv,review}\,:

\begin{equation}
\rho_{curv} = \frac{1}{f^{\prime}(R)} \left \{ \frac{1}{2} \left [ f(R) - R f^{\prime}(R) \right ]
- 3 H \dot{R} f^{\prime \prime}(R) \right \} \ ,
\label{eq: rhocurv}
\end{equation}

\begin{eqnarray}
p_{curv} & = \displaystyle{\frac{1}{f^{\prime}(R)}} & \left \{ (2 H \dot{R} + \ddot{R}) f^{\prime \prime}(R) +
\dot{R}^2 f^{\prime \prime \prime}(R) \right . \nonumber \\
~ & ~ & \ \ \left . + \frac{1}{2} \left [ f(R) - R f^{\prime}(R) \right ] \right \}  \ .
\label{eq: pcurv}
\end{eqnarray}
This approach is particularly useful since it allows to interpret the non-Einstein part of  gravitational interaction as a {\it new fluid} with energy density and pressure given by (\ref{eq: rhocurv}) and (\ref{eq: pcurv}) respectively. The barotropic factor for such a fluid turns out to be\,:

\begin{equation}
w_{curv} = -1 + \frac{f^{\prime \prime}(R) \ddot{R} + \left [ f^{\prime \prime \prime}(R) \dot{R} - H f^{\prime \prime}(R) \right ]}
{\left [ f(R) - R f^{\prime}(R) \right ]/2 - 3 H f^{\prime \prime}(R)} \ .
\label{eq: wcurv}
\end{equation}
A leading role is played by the form chosen for $f(R)$. Following  \cite{capozcurv}, we set $f(R) = f_0 R^n$ with $f_0$ constant. Unfortunately, it is not possible to analitically solve the Friedmann equations for the model with both matter and curvature contributions. However, if we neglect the matter term\footnote{The model we obtain by this ansatz should be considered as a toy model that is able to give some useful insights on the true model. However, it is worth noting that the matter contribution is indeed much smaller than the curvature one and it should be indeed negligible if only the baryonic matter is considered provided that the theory is able to erase the need for dark matter.} (i.e. we set $\rho_M = 0$), it is possible to find power law solutions of the Friedmann equations as $a(t) = (t/t_0)^{\alpha}$ with \cite{review}\,:

\begin{equation}
\alpha = \frac{2 n^2 - 3 n +1}{2 - n} \ .
\label{eq: alphancurv}
\end{equation}
This gives an entire class\footnote{Actually, there is another class of models characterized by $\alpha = 0$ and $n = 0, 1/2, 1$, but it is ruled out by the non vanishing value of the Hubble constant.} of cosmological models with constant deceleration parameter\,:

\begin{equation}
q(t) = q_0 = \frac{1 - \alpha}{\alpha} = - \frac{2 n^2 - 2 n - 1}{2 n^2 - 3 n + 1} \ .
\label{eq: qzcurv}
\end{equation}
In order to get an accelerated expansion ($\alpha > 0$ and $q_0 < 0$), the parameter $n$ has to satisfy the following constraint\,:

\begin{equation}
n \in \left [ -\infty, \frac{1 - \sqrt{3}}{2} \right ] \cup \left [ \frac{1 + \sqrt{3}}{2}, \infty \right ] \ .
\label{eq: nranges}
\end{equation}
We will refer to models with $n$ in the first (second) range as {\it CurvDown} ({\it CurvUp}) models respectively and use the method described in the following section to constrain the two parameters $(n, H_0)$ which completely assign the model.

\section{The method}

Most of the tests recently used to constrain cosmological parameters (such as the SNeIa Hubble diagram and the angular size\,-\,redshift) are essentially distance\,-\,based methods. The proposal of Dalal et al. \cite{DAJM01} to use the lookback time to high redshift objects is thus particularly interesting since it relies on a completely different observable. The lookback time is defined as the difference between the present day age of the universe and its age at redshift $z$ and may be computed as\,:

\begin{equation}
t_L(z, {\bf p}) = t_H \int_{0}^{z}{\frac{dz'}{(1 + z') E(z', {\bf p})}}
\label{eq: deftl}
\end{equation}
where $t_H = 1/H_0 = 9.78 h^{-1} \ {\rm Gyr}$ is the Hubble time (with $h$ the Hubble constant in units of $100 \ {\rm km \ s^{-1} \ Mpc^{-1}}$), $E(z, {\bf p}) = H(z)/H_0$ is the dimensionless Hubble parameter and we have denoted with $\{{\bf p}\}$ a set of parameters characterizing a given cosmological model. It is worth noting that, by definition, the lookback time is not sensible to the present day age of the universe $t_0$ so that it is (at least in principle) possible that a model fits well the data on the lookback time, but nonetheless it predicts a completely wrong value for $t_0$. This latter parameter can be evaluated from Eq.(\ref{eq: deftl}) by simply changing the upper integration limit from $z$ to infinity. This shows that it is indeed a different quantity since it depends on the full evolution of the universe and not only on how the universe evolves from the redshift $z$ to now. That is why we will explicitly introduce this quantity as a further constraint.

Let us now discuss how we use the lookback time and the age of the universe to test a given cosmological model. To this end, let us consider an object $i$ at redshift $z$ and denote by $t_i(z)$ its age defined as the difference between the age of the universe when the object was born, i.e. at the formation redshift $z_F$, and the one at $z$. It is\,:

\begin{eqnarray}
t_i(z) & = & \displaystyle{\int_{z}^{\infty}{\frac{dz'}{(1 + z') E(z', {\bf p})}} - \int_{z_F}^{\infty}{\frac{dz'}{(1 + z') E(z', {\bf p})}}} \nonumber \\
~ & = & \displaystyle{\int_{z}^{z_F}{\frac{dz'}{(1 + z') E(z', {\bf p})}}} \nonumber \\
~ & = & t_L(z_F) - t_L(z) \ .
\label{eq: titl}
\end{eqnarray}
where, in the last row, we have used the definition (\ref {eq: deftl}) of the lookback time. Suppose now we have $N$ objects and we have  been able to estimate the age $t_i$ of the object at redshift $z_i$ for $i = 1, 2, \ldots, N$. Using the previous relation, we can estimate the lookback time $t_{L}^{obs}(z_i)$ as\,:

\begin{eqnarray}
t_{L}^{obs}(z_i) & = & t_L(z_F) - t_i(z) \nonumber \\
~ & = & [t_0^{obs} - t_i(z)] - [t_0^{obs} - t_L(z_F)] \nonumber \\
~ & = & t_{0}^{obs} - t_i(z) - df \ ,
\label{eq: deftlobs}
\end{eqnarray}
where $t_{0}^{obs}$ is the today estimated age of the universe and a {\it delay factor} can be defined as\,:

\begin{equation}
df = t_0^{obs} - t_L(z_F) \ .
\end{equation}
The delay factor is introduced to take into account our ignorance of the formation redshift $z_F$ of the object. Actually, what can be measured is the age $t_i(z)$ of the object at redshift $z$. To estimate $z_F$, one should use Eq.(\ref{eq: titl}) assuming a background cosmological model. Since our aim is to determine what is the background cosmological model, it is clear that we cannot infer $z_F$ from the measured age so that this quantity is completely undetermined. It is worth stressing that, in principle, $df$ should be different for each object in the sample unless there is a theoretical reason to assume the same redshift at the formation of all the objects. If this is indeed the case (as we will assume later), then it is computationally convenient to consider $df$ rather than $z_F$ as the unknown parameter to be determined from the data. We may then define a likelihood function as\,:

\begin{equation}
{\cal{L}}_{lt}({\bf p}, df) \propto \exp{[-\chi^{2}_{lt}({\bf p}, df)/2]}
\label{eq: deflikelt}
\end{equation}
with\,:

\begin{displaymath}
\chi^{2}_{lt} = \displaystyle{\frac{1}{N - N_p + 1}}
\left \{ \left [ \frac{t_{0}^{theor}({\bf p}) - t_{0}^{obs}}{\sigma_{t_{0}^{obs}}} \right ]^2 \right .
\end{displaymath}
\begin{equation}
\ \ \ \ \ \ \ \ \ \ \ \ \ \ \ \ \ \ \ \ \ \ \ \ \
\left . + \sum_{i = 1}^{N}{\left [ \frac{t_{L}^{theor}(z_i, {\bf p}) -
t_{L}^{obs}(z_i)}{\sqrt{\sigma_{i}^2 + \sigma_{t}^{2}}} \right ]^2} \right \}
\label{eq: defchi}
\end{equation}
where $N_p$ is the number of parameters of the model, $\sigma_t$ is the uncertainty on $t_{0}^{obs}$, $\sigma_i$ the one on $t_{L}^{obs}(z_i)$ and the superscript {\it theor} denotes the predicted values of a given quantity. Note that the delay factor enters the defintion of $\chi^2_{lt}$ since it determines $t_{L}^{obs}(z_i)$ from $t_i(z)$ in virtue of Eq.(\ref{eq: deftlobs}), but the theoretical lookback time does not depend on $df$.

In principle, such a method should work efficiently to discriminate among the various dark energy models. Actually, this is not exactly the case due to the paucity of the available data which leads to large uncertainties on the estimated parameters. In order to partially alleviate this problem, it is convenient to add further constraints on the model by using a Gaussian prior\footnote{The need for a prior to reduce the parameter uncertainties has been also advocated in previous works using the age of old high redshift galaxies as a cosmological test. For instance, in Ref.\,\cite{LA00} a strong prior on $\Omega_M$ is introduced to better constrain the dark energy equation of state. It is likely that extending the dataset to higher redshifts and reducing the uncertainties on the age estimate will allow to avoid resorting to priors.} on the Hubble constant, i.e. redefining the likelihood function as\,:

\begin{equation}
{\cal{L}}({\bf p}) \propto {\cal{L}}_{lt}({\bf p})
\exp{\left [ -\frac{1}{2} \left ( \frac{h - h^{obs}}{\sigma_h} \right )^2 \right ]} \propto \exp{[- \chi^2({\bf p})/2]}
\label{eq: deflike}
\end{equation}
where we have absorbed $df$ in the set of parameters ${\bf p}$ and have defined\,:
\begin{equation}
\chi^2 = \chi^{2}_{lt} + \left ( \frac{h - h^{obs}}{\sigma_h} \right )^2
\label{eq: newchi}
\end{equation}
with $h^{obs}$ the estimated value of $h$ and  $\sigma_h$ its uncertainty. We use the HST Key project results \cite{Freedman} setting $(h, \sigma_h) = (0.72, 0.08)$. Note that this estimate is independent of the cosmological model since it has been obtained from local distance ladder methods.

The best fit model parameters ${\bf p}$ may  be obtained by maximizing ${\cal{L}}({\bf p})$ which is equivalent to minimize the $\chi^2$ defined in Eq.(\ref{eq: newchi}). It is worth stressing that such a function should not be considered as a {\it statistical $\chi^2$} in the sense that it is not forced to be of order 1 for the best fit model to consider a fit as successful. Actually, such an interpretation is not possible since the errors on the measured quantities (both $t_i$ and $t_0$) are not Gaussian distributed and, moreover, there are uncontrolled systematic uncertainties that may also dominate the error budget. Nonetheless, a qualitative comparison among different models may be obtained by comparing the values of this pseudo $\chi^2$ even if this should not be considered as a definitive evidence against a given model.

Having more than one parameter, we obtain the best fit value of each single parameter $p_i$ as the value which maximizes the marginalized likelihood for that parameter defined as\,:

\begin{equation}
{\cal{L}}_{p_i} \propto \int{dp_1 \ldots
\int{dp_{i - 1} \int{dp_{i + 1} \ldots \int{dp_n \ {\cal{L}}({\bf p})}}}} \ .
\label{eq: deflikemar}
\end{equation}
After having normalized  the marginalized likelihood to 1 at maximum, we compute the $1 \sigma$ and $2 \sigma$ confidence limits (CL) on that parameter by solving ${\cal{L}}_{p_i} = \exp{(-0.5)}$ and ${\cal{L}}_{p_i} = \exp{(-2)}$ respectively.

\section{The data}

In order to apply the method outlined above, we need a set of distant objects whose age can be somehow estimated. Clusters of galaxies seem to be ideal candidates in this sense since they can be detected up to high redshift and their redshift at formation epoch \footnote{It is worth stressing that, in literature, the cluster formation redshift is defined as the redshift at which the last episode of star formation happened. In this sense, we should modify our definition of $df$ by adding a constant term which takes care of how long is the star formation process and what is the time elapsed from the beginning of the universe to the birth of the first cluster of galaxies. For this reason, it is still possible to consider the delay factor to be the same for all clusters, but it is not possible to infer $z_F$ from the fitted value of $df$ because we do not know the detail of the star formation history. We stress that this approach is particular useful since it allows us to overcome the problem to consider lower limits of the universe age at $z$ rather than  the actual values.} is almost the same for all the clusters. Furthermore, it is relatively easy to estimate their age from photometric data only.

To this end, the color of their component galaxies, in particular the reddest ones, is all what is needed. Actually, the stellar populations of the reddest galaxies become redder and redder as they evolve. It is just a matter, then, to assume a stellar population synthesis model, and to look at how old the latest episode of star formation should be happened in the galaxy past to produce colors as red as the observed ones. This is what we will refer to as {\it color age}. The main limitation of the method relies in the stellar population synthesis model, and on a few (unknown) ingredients (among which the metallicity and the star formation rate law).

The choice of the evolutionary model is a key step in the estimate of the color age and the main source of uncertainty \cite{Worthey}. An alternative and more robust route to cluster age is to consider the color scatter (see \cite{Bower} for an early application of this approach). The argument, qualitatively, goes in this way\,: if galaxies have an extreme similarity in their color and nothing is conspiring to make the color scatter surreptitiously small, then the latest episode of star formation should happened in the galaxy far past, otherwise the observed color scatter would be larger. Quantitatively, the scatter in color should thus be equal to  the derivative of color with time multiplied the scatter of star formation times. The first quantity may be predicted using population synthesis models and turns out to be almost the same for all the evolutionary models thus significantly reducing the systematic uncertainty. We will refer to the age estimated by this method as {\it scatter age}.

The dataset we need to apply the method described in the previous section may now be obtained using the following procedure. First, for a given redshift $z_i$, we collect the colors of the redddest galaxies in a cluster at that redshift and then use one of the two methods outlined above to determine the color or the scatter age of the cluster. If more than one cluster is available at that redshift, we average the results from different clusters in order to reduce systematic errors. Having thus obtained $t_i(z_i)$, we then use Eq.(\ref{eq: deftlobs}) to estimate the value of the lookback time at that redshift. Actually, what we measure is $t_{L}^{obs}(z_i) + df$ that is the quantity that enters the definition (\ref{eq: defchi}) of $\chi^2_{lt}$ and then the likelihood function.

To estimate the color age, following \cite{Andreon3,Andreon4}, we have chosen, among the various available stellar population synthesis models, the Kodama and Arimoto one \cite{Kodama1}, which, unlike other models, allows a chemical evolution neglected elsewhere. This gives us three points on the diagram $z$ vs. $t_{L}^{obs}$ obtained by applying the method to a set of six clusters at three different redshifts as detailed in the left part of Table\,1. Using a large sample of low redshift SDSS clusters imaged, one of us \cite{Andreon1} has evaluated the scatter age for clusters age at $z = 0.10$ and $z = 0.25$, while Blakeslee et al. \cite{Blakeslee} applied the same method to a single, high redshift $(z = 1.27)$ cluster. Collecting the data using both  the color age and the scatter age, we end up with a sample of $\sim 160$ clusters at six redshifts (listed in Table\,1) which probe the redshift range $(0.10, 1.27)$. This nicely overlaps the one probed by SNeIa Hubble diagram so that a comparison among our results and those from SNeIa is possible. We assume a $\sigma = 1 \ {\rm Gyr}$ as uncertainty on the cluster age, no matter what is the method used to get that estimate. Note that this is a very conservative choice. Actually, if the error on the age were so large, the color\,-\,magitude relation for reddest cluster galaxies should have a large scatter that is not observed. We have, however, chosen such a large error to take qualitatively into account the systematic uncertainties related to the choice of the evolutionary model.

\begin{table}
\begin{center}
\begin{tabular}{|c|c|c|c|c|c|c|c|}
\hline
 \multicolumn{4}{|c|}{Color age} & \multicolumn{4}{|c|}{Scatter age} \\
\hline $z$  & N & Age (Gyr) & Ref & $z$  & N & Age (Gyr) & Ref \\
\hline\hline
0.60 & 1 & 4.53 & \cite{Andreon3} & 0.10 & 55 & 10.65 & \cite{Andreon1} \\
0.70 & 3 & 3.93 & \cite{Andreon4} & 0.25 & 103 & 8.89 & \cite{Andreon1} \\
0.80 & 2 & 3.41 & \cite{Andreon3} & 1.27 &   1 & 1.60 & \cite{Blakeslee} \\
\hline
\end{tabular}
\end{center}
\caption{Main properties of the cluster sample used  for the analysis. The data in the left part of the Table refers to clusters whose age has been estimated from the color of the reddest galaxies (color age), while that of clusters in the right part has been obtained by the color scatter (scatter age). For each data point, we give the redshift $z$, the number $N$ of clusters used, the age estimate and the relevant reference.}
\end{table}

Finally, we need an estimate of $t_{0}^{obs}$ to apply the method. Following Rebolo et al. \cite{VSA}, we choose $(t_{0}^{obs}, \sigma_t) = (14.4, 1.4) \ {\rm Gyr}$ as obtained by a combined analysis of the WMAP and VSA data on the  CMBR anisotropy spectrum and SDSS galaxy clustering. Actually, this estimate is model dependent since Rebolo et al. \cite{VSA} implicitly assumes that the $\Lambda$CDM model is the correct one. However, this value is in perfect agreement with $t_{0}^{obs} = 12.6^{+3.4}_{-2.4} \ {\rm Gyr}$ determined from globular clusters age \cite{Krauss} and $t_{0}^{obs} > 12.5 {\pm} 3.5 \ {\rm Gyr}$ from radioisotopes studies \cite{Cayrel}. For this reason, we are confident that no systematic error is induced on our method using the Rebolo et al. estimate for $t_{0}^{obs}$ even when testing cosmological models
other than the $\Lambda$CDM one.

\section{Results}

We have applied the method  outlined above to the dark energy models described in Sect.\,II in order to constrain their parameters and see if they are viable candidates to explain cosmic acceleration. To this aim, let us note that each one of the presented models is fully described by only few parameters which are\footnote{We drop the subscript $0$ from $\Omega_{M,0}$ since it does not give rise to any confusion here. Also, we use the dimensionless parameter $h$ instead of the Hubble constant $H_0$.} $(\Omega_M, h, w)$ for the QCDM model, $(q_0, h)$ for the parametric density model and $(n, h)$ for the curvature quintessence. For all three models, there is still another parameter entering the fit, that is the delay factor $df$, which we will marginalize over since it is not interesting for our aims.

\begin{figure}
\centering
\resizebox{8.5cm}{!}{\includegraphics{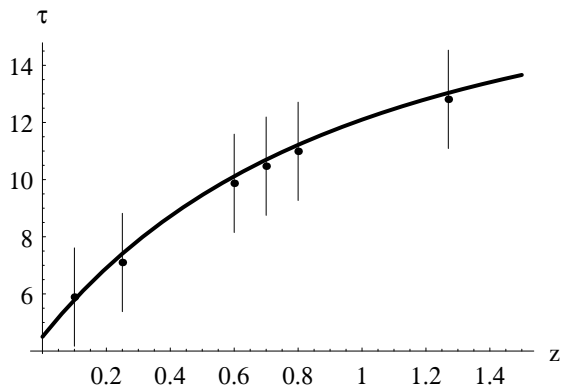}}
\hfill
\caption{Comparison among predicted and observed values of $\tau = t_L(z) + df$ for the best fit QCDM model.}
\label{fig: tauqcdm}
\end{figure}

\begin{figure}
\centering \resizebox{8.5cm}{!}{\includegraphics{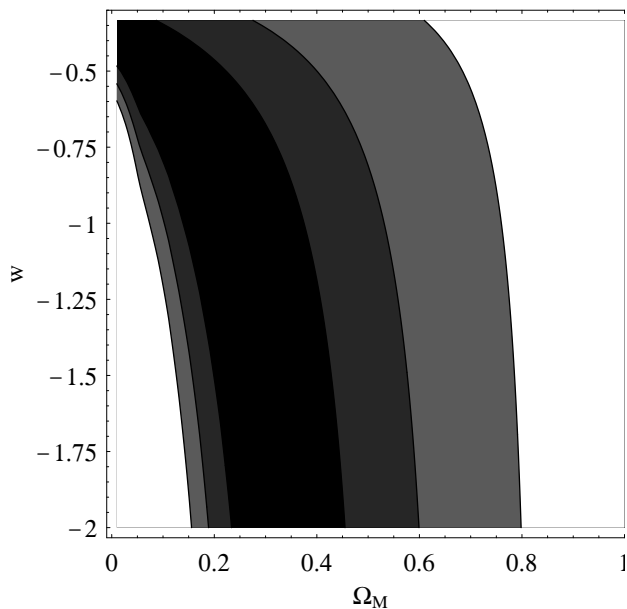}}
\hfill
\caption{The $1 \sigma$ and $2 \sigma$ confidence regions in the plane $(\Omega_M, w)$ for the QCDM model after marginalizing over the Hubble constant $h$ and the delay factor $df$.}
\label{fig: clqcdm}
\end{figure}

Let us first consider the QCDM model. The main results are plotted in Figs.\,\ref{fig: tauqcdm} and \ref{fig: clqcdm}. In the first plot, we compare the estimated clusters age with the quantity\,:

\begin{equation}
\tau(z) = t_L(z) + df
\label{eq: deftau}
\end{equation}
using the best fit values for the model parameters and the delay factor which turn out to be\,:

\begin{displaymath}
(\Omega_M, h, w) = (0.25, 0.70, -0.81) \ \ \ \ , \ \ \ \ df = 4.5 \ {\rm Gyr}
\end{displaymath}
giving $\chi^2 \simeq 0.04$. The $\chi^2$ value for the best fit parameters (both for the QCDM model and the other dark energy models considered) turns out to be quite small suggesting that we have seriously overestimated the errors. This is not surprising given the (somewhat arbitrary) way we have fixed the uncertainties on the estimated age of the clusters. That this is likely to be the case is also suggested by a qualitative argument. One could rescale the errors on $t_{L}^{obs}(z_i)$ in such a way that $\chi^2 = 1$ for the best fit model. Since for the best fit QCDM model, $\chi^2 \simeq \chi^2_{lt}$, this leads to multiply by almost $1/5$ the uncertainties on $t_{L}^{obs}(z_i)$. If the error on $t_0$ were negligible, this means that we should reduce the uncertainty on the cluster age from 1 Gyr to 0.2 Gyr that is indeed a more realistic value. The presence of an error on $t_0^{obs}$ slightly complicates this argument, but does not change the main conclusion. We are thus confident that the very low value of the $\chi^2$ we get for the best fit model is only due to overestimating the uncertainties on the clusters ages. However, we do not perform any rescaling of the uncertainties since, to this end, we should select a priori a model as the confidence one which is contrary to the philosophy of the paper. It is worth stressing, however, that such rescaling does not affect anyway the main results.

\begin{figure}
\centering
\resizebox{8.5cm}{!}{\includegraphics{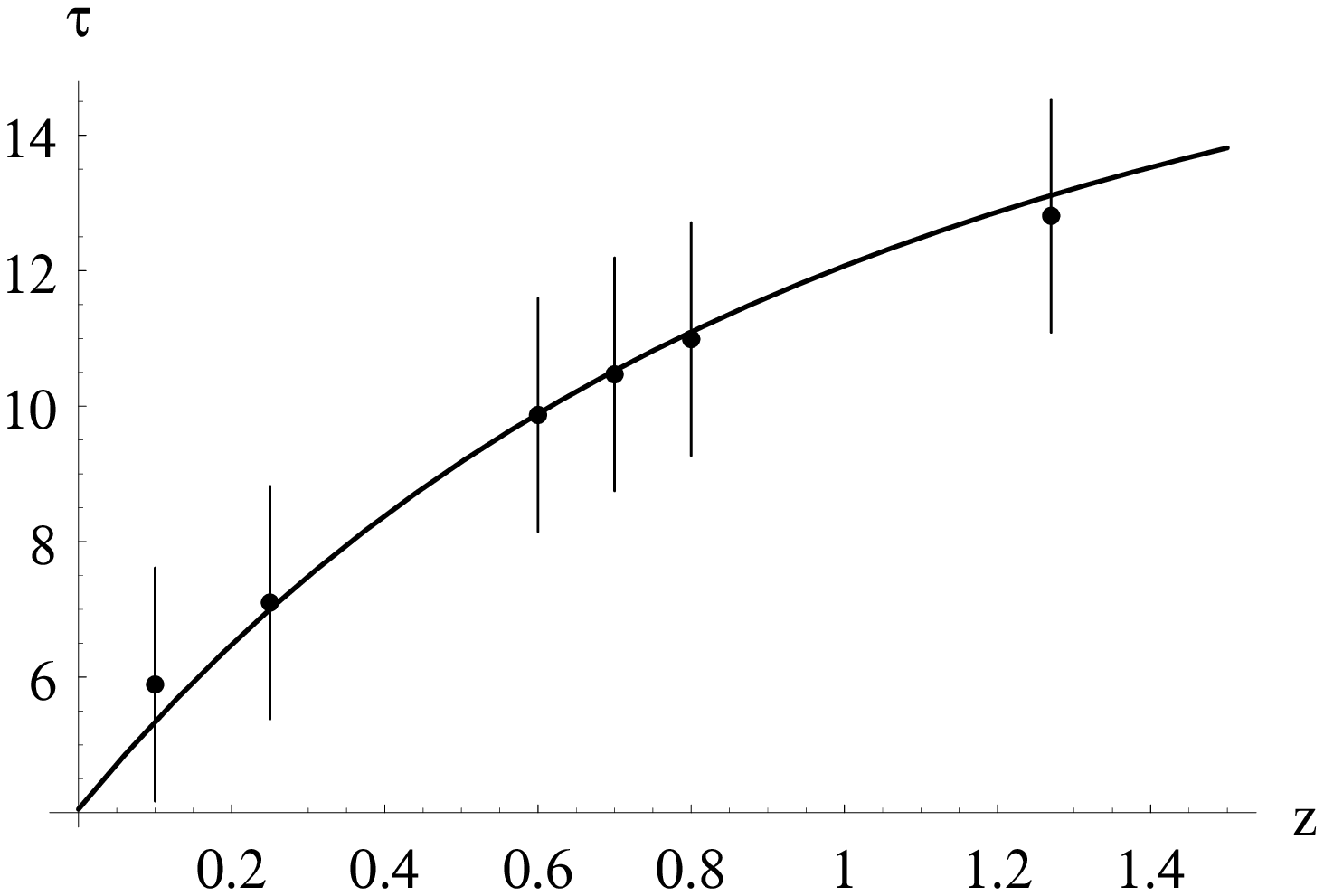}}
\hfill
\caption{Comparison among predicted and observed values of $\tau = t_L(z) + df$ for the best fit $\Lambda$CDM model.}
\label{fig: taulcdm}
\end{figure}

\begin{figure}
\centering \resizebox{8.5cm}{!}{\includegraphics{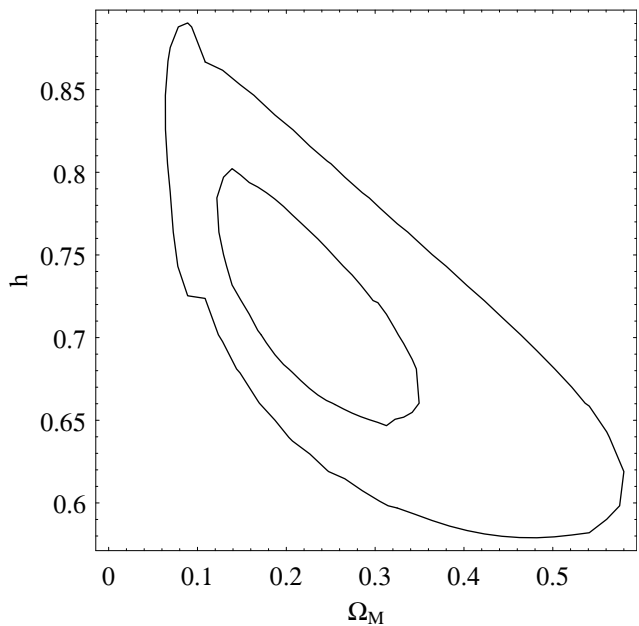}}
\hfill
\caption{The $1 \sigma$ and $2 \sigma$ confidence regions in the plane $(\Omega_M, h)$ for the $\Lambda$CDM model.}
\label{fig: cllcdm}
\end{figure}

Fig.\,\ref{fig: clqcdm} shows the $1 \sigma$ and $2 \sigma$ confidence levels in the $(\Omega_M, w)$ plane obtained by marginalizing over the Hubble constant and the delay factor. Two interesting considerations may be drawn from that plot. First, we note that also {\it phantom} models (i.e. models with $w < -1$ violating the weak energy condition) are allowed by the data. This is in agreement with recent results coming from fitting the QCDM model to the SNeIa Hubble diagram and the CMBR anisotropy spectrum \cite{Phantom}. Unfortunately, a direct comparison is not possible since the marginalized likelihood is too flat to get any constraints on $w$ so that all the values in the range tested $(-2 \le w \le 1/3)$ are well within the $1 \sigma$ CL. Although essentially due to the paucity of the data, this result is also a consequence of having not used any prior on $\Omega_M$ as it is usually done in other analyses. By using the procedure described at the end of Sect.\,III, we obtain the following estimates for the other QCDM parameters\,:

\begin{displaymath}
\Omega_M \in (0.13, 0.39) \ \ \ \ , \ \ \ \ h \in (0.63, 0.77) \ \ \ \ (1 \sigma \ {\rm CL}) \ ,
\end{displaymath}

\begin{displaymath}
\Omega_M \in (0.01, 0.62) \ \ \ \ , \ \ \ \ h \in (0.56, 0.84) \ \ \ \ (2 \sigma \ {\rm CL}) \ .
\end{displaymath}
Given that we are able to give only weak constraints on the QCDM model, from now on we will only dedicate our attention to the case $w = -1$, i.e. to the $\Lambda$CDM model and do not discuss anymore the results for the QCDM model. The best fit parameters for the cosmological constant model turn out to be\,:

\begin{displaymath}
(\Omega_M, h) = (0.22, 0.71) \ \ \ \ , \ \ \ \ df = 4.05 \ {\rm Gyr}  \ \ \ \ (\chi^2 \simeq 0.09)
\end{displaymath}
that gives rise to the curve $\tau(z)$ shown in Fig.\,\ref{fig: taulcdm}, while Fig.\,\ref{fig: cllcdm} reports the confidence regions in the $(\Omega_M, h)$ plane after marginalizing over the delay factor. From the marginalized likelihood functions, we get\,:

\begin{displaymath}
\Omega_M \in (0.10, 0.35) \ \ \ \ , \ \ \ \ h \in (0.63, 0.78) \ \ \ \ (1 \sigma \ {\rm CL}) \ ,
\end{displaymath}

\begin{displaymath}
\Omega_M \in (0.06, 0.59) \ \ \ \ , \ \ \ \ h \in (0.56, 0.85) \ \ \ \ (2 \sigma \ {\rm CL}) \ .
\end{displaymath}

The $\Lambda$CDM model has been widely tested against  a large set of different astrophysical data. This offers us the possibility to cross check both the model and the validity of method. It is instructive, in this sense, to compare our results with those coming from the fit to the SNeIa Hubble diagram. For instance, Barris et al. \cite{Barris04} used a set of 120 SNeIa up to $z = 1.03$ finding $\Omega_M = 0.33$ as best fit value with a large uncertainty (not quoted explicitly, but easy to see in their Fig.12) in good agreement with our result. A more recent result has been obtained by Riess et al. \cite{Riess04} which have used a SNeIa Hubble diagram extending up to $z = 1.55$ and have found $\Omega_M = 0.29_{-0.03}^{+0.05}$ still in agreement with our result. The $\Lambda$CDM model has also been tested by means of the angular size\,-\,redshift relation. Using a catalog of ultracompact radio sources and taking carefully into account systematic uncertainties and selection effects, Jackson \cite{Jack03} has found $\Omega_M = 0.24_{-0.07}^{+0.09}$ in almost perfect agreement with our estimate.

\begin{figure}
\centering
\resizebox{8.5cm}{!}{\includegraphics{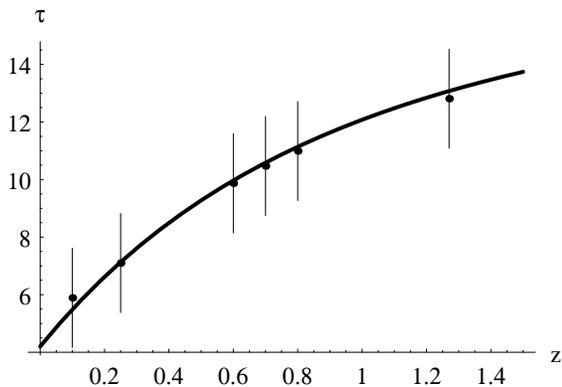}}
\hfill
\caption{Comparison among predicted and observed values of $\tau = t_L(z) + df$ for the best fit parametric density model.}
\label{fig: tauhobbit}
\end{figure}

Neither Barris et al. \cite{Barris04} nor Riess et al. \cite{Riess04} quote a best fit value for $h$ since this parameter is infinitely degenerate with the supernovae absolute magnitude $M$ when fitting to the SNeIa Hubble diagram. Nonetheless, SNeIa may still be used to determine $h$ by fitting the linear Hubble law to the low redshidt ($z < 0.1$) SNeIa. Using this method, Daly \& Djorgovski \cite{DD04} have found $h = 0.664 {\pm} 0.08$ in good agreement with our result. Moreover, it is worth noting that our estimate for $h$ turns out to be in agreement with estimates coming from different methods such as various local standard candles \cite{Freedman}, the Sunyaev\,-\,Zel'dovich effect in galaxy clusters \cite{SZ-Hz} and time delays in multiply imaged quasars \cite{Hz-lensing}. Finally, let us quote the results obtained by Tegmark et al. \cite{SDSS03} which have performed a combined fit of the $\Lambda$CDM model to both the WMAP data on the CMBR anisotropy spectrum and the galaxy power spectrum measured by more than 200,000 galaxies surveyed by the SDSS collaboration. They find $\Omega_M = 0.30 {\pm} 0.04$ and $h = 0.70^{+0.04}_{-0.03}$ in very good agreement with our results.

Perhaps, the most interesting result of testing the $\Lambda$CDM model with our method is not the success of this model (since it has yet been shown by a lot of evidences), but the substantial agreement among our estimates of the parameters $(\Omega_M, h)$ and those coming from different kinds of data. This is quite encouraging since it is an important successful cross check and makes us confident about the results one could obtain by applying it to other cosmological models.

\begin{figure}
\centering \resizebox{8.5cm}{!}{\includegraphics{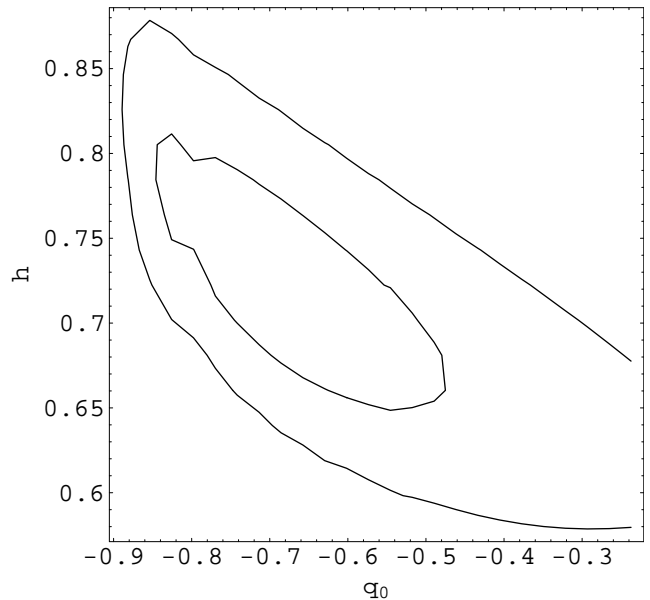}}
\hfill \caption{The $1 \sigma$ and $2 \sigma$ confidence regions in the plane $(q_0, h)$ for the parametric density model.}
\label{fig: clhobbit}
\end{figure}

Let us now examine the results obtained for the parametric density model, pictorially shown in Figs.\,\ref{fig: tauhobbit} and \ref{fig: clhobbit}. The best fit model is obtained for the following values of the parameters $(q_0, h)$ and of the delay factor\,:

\begin{displaymath}
(q_0, h) = (-0.68, 0.71) \ \ \ \ , \ \ \ \ df = 4.20 \ {\rm Gyr}  \ \ \ \ (\chi^2 \simeq 0.07) \ .
\end{displaymath}
Marginalizing over $df$, we get\,:

\begin{displaymath}
q_0 \in (-0.81, -0.47) \ \ \ \ , \ \ \ \ h \in (0.64, 0.78) \ \ \ \ (1 \sigma \ {\rm CL}) \ ,
\end{displaymath}

\begin{displaymath}
q_0 \in (-0.89, -0.24) \ \ \ \ , \ \ \ \ h \in (0.58, 0.85) \ \ \ \ (2 \sigma \ {\rm CL}) \ .
\end{displaymath}
Note that the $2 \sigma \ {\rm CL}$ on the $q_0$ parameter  has been truncated at the upper end since it formally extends to values higher than the physically acceptable one.

In \cite{Hobbit}, the parameters of this model have been constrained by using both the SNeIa Hubble diagram and the angular size\,-\,redshift relation. In particular, fitting the model to the SNeIa Hubble diagram gives $h = 0.64_{-0.05}^{+0.08}$, while  the physically acceptable range for $q_0$ turns out to be in agreement with the data for $q_0 = -0.42$ as best fit value. The present day deceleration parameter $q_0$ is better constrained using the data listed in Jackson \cite{Jack03} to perform the angular size\,-\,redshift test thus obtaining $q_0 = -0.64_{-0.12}^{+0.10}$ \cite{Hobbit}. Both these results are in very good agreement with our estimates so that we conclude that the parametric density model is  a viable candidate alternative to the $\Lambda$CDM model.

\begin{figure}
\centering \resizebox{8.5cm}{!}{\includegraphics{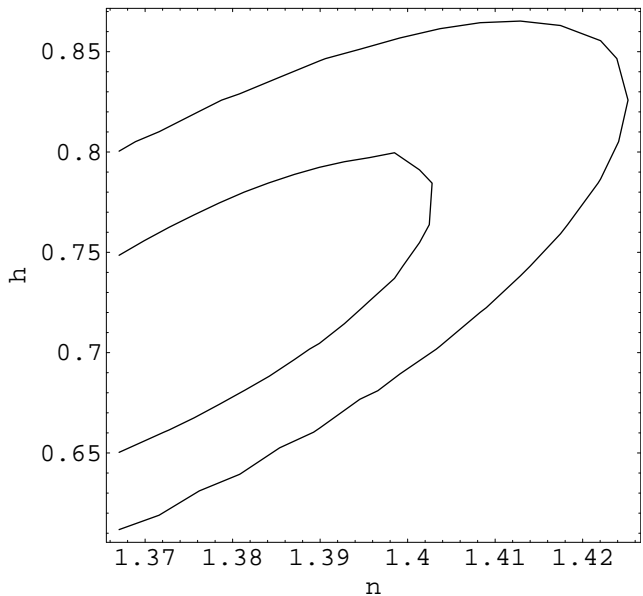}}
\hfill
\caption{The $1 \sigma$ and $2 \sigma$ confidence regions  in the plane $(n, h)$ for curvature quintessence in the CurvUp regime.}
\label{fig: clcurvup}
\end{figure}

Finally, let us discuss the results  for the curvature quintessence. Figs.\,\ref{fig: clcurvup} and \ref{fig: clcurvdown} report the confidence regions in the plane $(n, h)$ for the CurvUp and CurvDown regime respectively after marginalizing over the delay factor. The first striking feature is that the contour plots are not closed so that the marginalized likelihood function gives only an upper (lower) limit to the parameter $n$ in the CurvUp (CurvDown) regime. Formally, we get the following estimates for the best fit values in the CurvUp and CurvDown regime respectively\,:

\begin{displaymath}
(n, h) = (1.367, 0.71) \ \ \ \ , \ \ \ \ df = 4.80 \ {\rm Gyr}  \ \ \ \ (\chi^2 \simeq 0.23) \ ,
\end{displaymath}

\begin{displaymath}
(n, h) = (-0.367, 0.74) \ \ \ \ , \ \ \ \ df = 4.80 \ {\rm Gyr}  \ \ \ \ (\chi^2 \simeq 0.21) \ ,
\end{displaymath}
but the best fit value for $n$ actually  lies outside the investigated range for this parameter. Being the confidence regions open, it is meaningless to give constraints on $h$, but nonetheless it is possible to infer the following limits\,:

\begin{figure}
\centering
\resizebox{8.5cm}{!}{\includegraphics{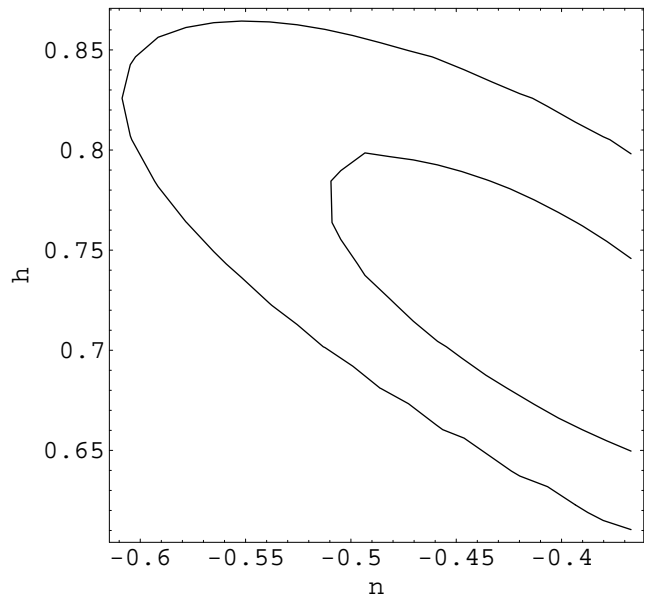}}
\hfill
\caption{The $1 \sigma$ and $2 \sigma$ confidence regions in the plane $(n, h)$ for curvature quintessence in the CurvDown regime.}
\label{fig: clcurvdown}
\end{figure}

\begin{displaymath}
n \le 1.402 \ \ {\rm at \ 1 \sigma \ CL} \ \ , \ \ n \le 1.424 \ \ {\rm at \ 2 \sigma \ CL} \ \ ,
\end{displaymath}

\begin{displaymath}
n \ge -0.508 \ \ {\rm at \ 1 \sigma \ CL} \ \ , \ \ n \ge -0.606 \ \ {\rm at \ 2 \sigma \ CL} \ \ ,
\end{displaymath}
where the first row refers to the CurvUp regime and  the second to the CurvDown one. These limits do not contradict the ranges determined by fitting to the SNeIa Hubble diagram \cite{curvfit}, but seems to be quite non-realistic. Actually, the deceleration parameter corresponding to these values of $n$ is quite small $(q_0 \sim -0.01)$ thus contradicting the evidence in favor of an accelerating universe. Moreover, the results in \cite{curvfit} have been obtained by using an old sample of SNeIa, including some SNeIa that have now been discarded from the Gold set in Riess et al. \cite{Riess04}. On the other hand, it is worth noting that fitting a power law scale factor to the angular size\,-\,redshift relation for compact radio sources gives $\alpha \simeq 1$ \cite{JDA03} which, by using Eq.(\ref{eq: alphancurv}), translates in an estimate for $n$ that is in good agreement with our result.

\section{Conclusions}

The impressive amount of data  indicating a spatially flat universe in accelerated expansion has posed the problem of dark energy and stimulated the search for cosmological models able to explain such unexpected behavior. Many rival theories have been proposed to solve the puzzle of the nature of  dark energy ranging from a rolling scalar field to a unified picture where a single exotic fluid accounts for the whole dark sector (dark matter and dark energy). Moreover, modifications of the gravity Lagrangian has also been advocated. Although deeply different in their underlying physics, all these scenarios share the common feature of well reproducing the available astrophysical data. It is worth stressing, however, that the most widely used cosmological tests and in particular the SNeIa Hubble diagram and the angular size\,-\,redhisft relation) are essentially based on distance measurements to high redshift objects and are thus affected by similar systematic errors. It is hence particularly interesting to look for  methods which are related to the estimates of different quantities. Being affected by other kinds of observational problems, such  methods could be considered as   cross checks for the results obtained by the usual tests and they should represent complementary probes for cosmological models.

The technique we have devised here is a first  step in this direction. We have used the present age of the universe and the lookback time to galaxy clusters to build a sort of {\it time diagram} of the universe in order to reconstruct its age evolution. Relying on stellar evolutionary codes, the estimate of the lookback time is related to a different astrophysics than the distance based methods and it is thus free from any problem connected with the evolution of standard candles (such as the SNeIa absolute magnitude and the intrinsic linear size of radio sources). Actually, this technique could be affected by its own systematics (such as, for instance, the degeneracy between age and metallicity), but these may be better controlled. Moreover, comparing the results thus obtained with those coming from distance based methods allows to strengthen the conclusions suggested by both techniques.

Motivated by these considerations,  we have first applied our method to the $\Lambda$CDM model in order to estimate the present day values of the matter density parameter $\Omega_M$ and of the dimensionless Hubble constant $h$ obtaining (at the $1 \sigma$ CL)\,:

\begin{displaymath}
\Omega_M = 0.22_{-0.10}^{+0.13} \ \ \ \ ,\ \ \ \ h = 0.71_{-0.08}^{+0.07} \ .
\end{displaymath}
These values are in agreement with the previous estimates in literature. This can be considered an independent confirmation not only of the viability of the $\Lambda$CDM model, but also of our method. It is worth noting that the $\Lambda$CDM scenario receives further support from this test and henceforth the cosmological constant $\Lambda$ still remains the best candidate to explain the dark energy puzzle from an observational point of view.

Nonetheless, the $\Lambda$CDM model is severely affected by conceptual problems so that it is worthwhile to look fo alternative approaches. This has stimulated a plenty of models where the cosmic acceleration is due to a dominant scalar field rolling down its potential. However, such a scheme, on the one hand, still does not solve the coincidence problem and, on the other hand, is plagued by the unidentified nature of the scalar field itself and the ignorance of its self interaction potential. These considerations have opened the way to different models that are able to give an accelerated expansion without the need of scalar fields. Two of these approaches have been tested in this paper.

In \cite{Hobbit}, it has been proposed a phenomenological unified model where a single fluid with a given energy density assigned by few parameters is able to fit both the SNeIa Hubble diagram and the angular size\,-\,redshift relation for ultracompact radio sources without the need of any cosmological constant or scalar field. This parametric density model has been tested here with lookback time method and the following estimates (at the $1 \sigma$ CL) for its characterizing parameters has been obtained\,:

\begin{displaymath}
q_0 = -0.68_{-0.13}^{+0.21} \ \ \ \ , \ \ \ \ h = 0.71 {\pm} 0.07 \ ,
\end{displaymath}
in agreement with the results from the test performed in \cite{Hobbit}. It is worth stressing that our method turns out to be more efficient than the usual fit to the SNeIa Hubble diagram for the parametric density model. It is also interesting to note that both $q_0$ and $t_0$ predicted by this model are in almost perfect agreement with those computed for the $\Lambda$CDM model with the parameters $(\Omega_M, h)$ as discussed above. This is not surprising since Eq.(\ref{eq: rhor}) shows that, nowadays, the energy density of the parametric density model is very similar to that of the $\Lambda$CDM model so that these two scenarios share most of the observable properties referred to today quantities. However, this does by no way mean that the two models are the same. Actually, it is the underlying philosophy that is completely different having now a single fluid rather than a cosmological constant dominating over the matter term. Both models predict similar  values ofthe today observed quantities  simply because they are tied to reproduce the same data and not because they are two different manifestations of the same underlying physics.

Another possible approach to the cosmic acceleration is to consider this feature as the first signal of the breakdown of the Einstein General Relativity at some characteristic scale. In this picture, the universe is still dominated by  standard matter, but the Friedmann equations have to be modified as consequence of a different gravity Lagrangian. This philosophy inspired  curvature quintessence scenarios where an effective dark energy is related to the properties of the function $f(R)$ which replaces the Ricci scalar $R$ in the gravity Lagrangian \cite{capozcurv,curvature,review}. The coupling between matter and curvature for a theory with $f(R) = f_0 R^n$ gives rise to fourth order nonlinear differential equations for the evolution of the scale factor that is not analytically solvable. This difficulty disappears if one considers a toy model with $\Omega_M = 0$ so that power law solutions, $a \sim t^{\alpha}$, are possible provided that $\alpha$ is linked to $n$ by Eq.(\ref{eq: alphancurv}). Such a model is particularly attractive from a theoretical viewpoint (since it allows to give a purely geometric interpretation of the dark energy) and has also been shown to fit well SNeIa Hubble diagram \cite{curvfit}. We have shown here that, while it is able to pass successfully the loockback time test implemented in this paper, only weak constraints can be imposed on the parameter $n$ with the best fit value lying in a region corresponding to decelerating rather than accelerating models. We are thus tempted to conclude that this scenario could be rejected with some confidence. Actually, this conclusion is also suggested by the recent result by Riess et al. \cite{Riess04} which, using the SNeIa Hubble diagram up to $z = 1.55$, have convincingly detected a change in the sign of the deceleration parameter $q$. The power law solutions predict, on contrary, a constant $q$ so that they are ruled out by the result in Riess et al. \cite{Riess04}. It is important to stress, however, that this is not a definitive exclusion of the curvature quintessence scenario. Actually, we have only considered power law solutions in absence of matter. It is indeed conceivable that including a matter term completely change the evolution of the scale factor possibly giving a past deceleration followed by the present acceleration thus leading the curvature quintessence model in agreement with what is suggested by the extended SNeIa Hubble diagram (see also \cite{curvature} and references therein for a further discussion of curvature models).

Having tested three different models, it is worth asking what is the better one. Unfortunately, this is not possible on the basis of the test results only. Comparing the $\chi^2$ values, one could naively conclude that the parametric density model is the best one since it gives the lowest $\chi^2$. However, while the difference among the $\chi^2$ values is significant between the parametric density model and the curvature quintessence scenario (almost an order of magnitude), it is too small (0.07 vs. 0.09) to conclude that the $\Lambda$CDM model is disfavored. We have thus to conclude that this test alone is not able to discriminate among these two dark energy candidates. On the other hand, Sandvik et al. \cite{Sandvik} recently claimed that UDE models are not viable because the growth of density perturbations will lead to matter power spectrum in disagreement with what is observed. This should be an evidence against the parametric density model. However, it is worth noting that Sandvik et al. explicitely consider the generalized Chaplygin gas model which is characterized by a negative squared sound speed that seems to be the main cause of the anomalous growth of perturbations. For the parametric density model, the sound speed is always positive definite so that it is likely that the argument of Sandvik et al. should be at least reconsidered.

Finally, we would briefly comment on the possibility to ameliorate our method by increasing the maximum redshift probed. To this aim, galaxy clusters do not appear as good candidates since it is quite difficult to detect a significative number of member galaxies up to redshifts larger than $\sim 1.3$. However, high redshift galaxies may be taken into account provided that they are detected in as more photometric bands as possible. This latter requirement is fundamental since it allows not only to better estimate the age of the galaxy, but also to infer constraints on its formation redshift $z_F$ that cannot be assumed to be the same for all the galaxies (as it has been possible for clusters). The $i$ drop out technique allows to discover galaxies up to redshift $z \sim 6$ \cite{Dick04} and thus let us hope to measure lookback time up to such high redshift. With its ability of both furnishing multicolor photometry of high redshift galaxies (and thus better estimates of their color ages and formation redshift) and increasing the number of SNeIa, the GOODS survey \cite{GOODS} seems to be the most promising source of cosmological constraints in the near future.

\acknowledgments{We warmly thank an anonymous referee for his comments that have helped us to improve the presentation.}

\end{document}